\documentclass[journal=jacsat,manuscript=article]{achemso}

\usepackage[T1]{fontenc}
\usepackage{graphicx}
\usepackage[version=3]{mhchem} 
\usepackage{dcolumn}
\usepackage{siunitx}
\DeclareSIUnit{\barpressure}{bar}
\DeclareSIUnit{\langmuir}{L}
\DeclareSIUnit{\photons}{photons}
\DeclareSIUnit\angstrom{\protect \text {Å}}
\usepackage{amsmath}
\usepackage{xcolor}
\DeclareGraphicsExtensions{.pdf,.png}

\bibstyle{achemso-demo}


\title{Valence Modifications in Hygroscopic VI\textsubscript{3} Degraded Crystals via Soft X-Ray Synchrotron Radiation}

\author{A. De Vita}
\email{alessandro.de.vita@tu-berlin.de}%
\affiliation{Institut für Physik und Astronomie, Technische Universität Berlin, Strasse des 17 Juni 135, 10623 Berlin, Germany\looseness=-1}%
\alsoaffiliation{Fritz Haber Institute of the Max Planck Society, Faradayweg 4--6, 14195 Berlin, Germany\looseness=-1}%
\author{V. Polewczyk}
\affiliation{Université Paris-Saclay, UVSQ, CNRS, GEMaC, 78000, Versailles, France\looseness=-1}%
\author{G. Panaccione}%
\affiliation{CNR - Istituto Officina dei Materiali (IOM),  S.S. 14, km 163.5, 34149 Trieste, Italy\looseness=-1}%
\author{G. Vinai}
\email{vinai@iom.cnr.it}
\affiliation{CNR - Istituto Officina dei Materiali (IOM),  S.S. 14, km 163.5, 34149 Trieste, Italy\looseness=-1}%

\begin{document}


\begin{abstract}

Among van der Waals crystals, transition metal trihalide VI\textsubscript{3} has driven attention for its magnetic and orbital properties. However, its chemical instability under ambient conditions make its exploitation challenging for technological implementation. In this context, here we show how synchrotron radiation soft X-rays \textcolor{black} {induce changes in the valence state of hygroscopic VI\textsubscript{3} crystals aged in high vacuum conditions.} 
By combining X-ray absorption and X-ray photoemission spectroscopies, we show as-cleaved and aged 
chemical degradation of VI\textsubscript{3} crystal surface, with the formation of vanadates, and its 
\textcolor{black} {valence modifications} under high-flux soft X-ray beam exposure, revealing that superficial hygroscopic contamination couples relatively weakly to the crystal surface.

\end{abstract}

\maketitle

\section{Introduction}

The family of two-dimensional (2D) materials has considerably expanded with the discovery of van der Waals crystals, with major consequences for the landscape of functional materials due to different properties emerging when moving from single crystals to monolayers or heterostructures \cite{Gibertini2019,Kajale2024}. Among them, triiodides XI\textsubscript{3} (X = Cr, V) stand out for their long-range magnetic properties and rich temperature-dependent phase diagram, with promising applications in the field of spin- and orbitronics \cite{Huang2017,Jiang2018a,Jiang2018b,Huang2018,Zhang2022,Fernandez2020}. In particular, VI\textsubscript{3} shows ferromagnetism down to monolayer thickness \cite{Lin2021} characterized by a reduced spin dimensionality \cite{DeVita2024}, a spin reorientation transition \cite{Hao2021}, and a large orbital magnetic moment at \SI{2}{\K} \cite{Hovancik2023} that gets quenched in proximity of the spin reorientation \cite{DeVita2024}. Conversely, it has been noted that XI\textsubscript{3} are highly unstable under ambient conditions, making their handling challenging. 
\textcolor{black}{The water solvation of VI\textsubscript{3} generates a variety of hydrated and solvated cationic vanadium ions, leading to a mixed valence state combining 3+, 4+ and 5+. In water solution, VI\textsubscript{3}  crystal structure is fully solvated by water molecules, leading to the formation of black droplets of V(H\textsubscript{2}O)\textsubscript{6}\cite{Krakowiak2012}, \textit{i.e.}} 
losing their crystallographic and magnetic properties, even in presence of protective oil layers \cite{Kratochvilova2022,Mastrippolito2022,Wang2023}. In this sense, VI\textsubscript{3} shares its hygroscopicity with CrI\textsubscript{3}
\textcolor{black}{. It has been shown that in the latter case light illumination can play a role in the formation of aquachromium iodides.}
\cite{Shcherbakov2018}. A road towards leveraging the hygroscopicity the same way as for more stable sister compounds, such as CrCl\textsubscript{3} \cite{Paolucci2024}, appears more challenging.

In this context, here we present a synchrotron radiation soft X‑ray investigation on the stability in ultra-high vacuum (UHV) conditions of VI\textsubscript{3} crystals upon aging. By combining absorption and photoemission spectroscopic measurements, we show that V 2\textit{p} edges  \textcolor{black}{present a modification of their valence due to interaction with the molecular contaminants present in the UHV chamber,} leading to the formation of \textcolor{black}{surface} vanadates. We then show how surface exposure to relatively mild photon fluxes is sufficient to alter the surface bonds of VI\textsubscript{3} with moisture \textcolor{black}{and oxygen, leading to a partial desorption of contaminants and reduction of vanadate species.}

\section{Experimental section}
\textbf{Sample \textcolor{black}{characteristics and} handling} -- Commercially available VI\textsubscript{3} crystals \textcolor{black}{(size $\approx\qtyproduct{3x3}{\mm}$)} stored in a glovebox filled with Ar atmosphere ($<\SI{0.5}{ppm}$ of O\textsubscript{2}, $<\SI{0.5}{ppm}$ of H\textsubscript{2}O) have been transferred in static atmosphere \textcolor{black}{via inert gas transfer shuttles}. \textcolor{black}{At high temperature, the crystal structure of VI\textsubscript{3} is monoclinic (space group $C2/m$, No. 12), with lattice parameters  $a=\SI{6.8416\pm0.0003}{\angstrom}$, $b=\SI{11.8387\pm0.0006}{\angstrom}$, and $c=\SI{6.9502\pm0.0004}{\angstrom}$ \cite{Tian2019}. The crystal structure and an image of a sample are displayed in Supporting Information, Fig.~S4}. A batch of samples (\textit{aged sample}) was cleaved inside the loadlock chamber (base pressure $<\SI{e-8}{\milli\barpressure}$), to expose the (0001) crystallographic plane of the clean surface, held in high vacuum (HV, pressure $\sim\SI{5e-9}{\milli\barpressure}$) for 72 hours, and measured afterwards. We estimate the exposure of the surface to be $\sim\SI{2600}{\langmuir}$. \textcolor{black}{Residual Gas Analysis (RGA) information on the composition of background gases is provided in the Supporting Information, Fig.~S3.} Another set of samples (\textit{as-cleaved sample}) was introduced from the glovebox, directly cleaved at the endstation chamber in ultrahigh vacuum (UHV, pressure $<\SI{3e-10}{\milli\barpressure}$), \textit{i.e.}, in better vacuum conditions (\textcolor{black}{surface exposure} $\sim\SI{2}{\langmuir}$), and immediately measured.

\textbf{XAS and XPS characterizations} -- X-ray absorption (XAS) and X-ray photoemission spectra (XPS) were carried out at APE-HE beamline of NFFA at the Elettra synchrotron radiation facility in Trieste \cite{Panaccione2009}. The sample surface was kept at \ang{45} with respect to the incident X-rays, for a beam footprint on the sample surface of around \qtyproduct{150x150}{\um}. XAS were measured in linear horizontal polarization at room temperature in total electron yield (TEY) mode, normalizing the intensity of sample current to the incident photon flux current at each energy value. The photon flux intensity was modulated by opening or closing mechanical exit slits placed after the spherical mirror, whose intensity is estimated by measuring the TEY signal of a gold-plated W grid placed in front of the endstation, and converting it into \textcolor{black}{\unit{\photons\per\s}} following beamline calibrations in ref.~\cite{Panaccione2009} . \textcolor{black}{The sample exposure time to X-rays is within the timeframe of the acquisition of the V \textit{L}\textsubscript{2,3} edge ($\approx$ 10 min)}. XPS measurements were taken with a ScientaOmicron R3000 analyzer, in horizontal polarization, with a photon energy of \SI{800}{\eV}, with the sample surface normal to the analyzer and at \ang{45} with respect to the X-ray beam.

\section{Results and discussion}

Firstly, we analyze the effects of surface aging in vacuum conditions on a VI\textsubscript{3} crystal cleaved at room temperature.

\begin{figure*}[htb]
\includegraphics[width=0.9\linewidth]{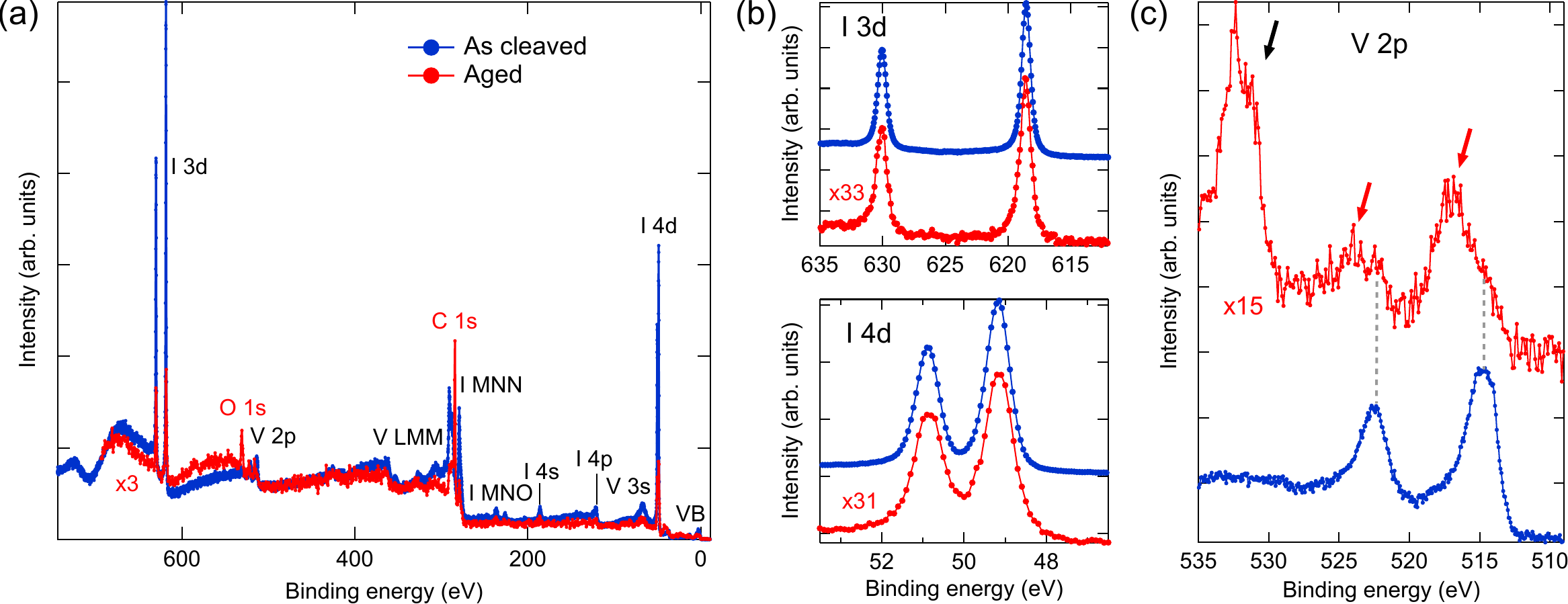}
\caption{\protect\label{fig:Fig1} (a) XPS survey at \SI{800}{\eV} on VI\textsubscript{3} as-cleaved (blue) and aged (red) samples, with element-specific peaks highlighted. (b) I 3\textit{d} and 4\textit{d} edges for as-cleaved (blue) and aged (red) samples. (c) V 2\textit{p} and O 1\textit{s} edges for as-cleaved (blue) and aged (red) samples. All measurements were done with a photon flux equivalent to $\SI{2.8e9}{\photons\per\s}$ on the mesh.}
\end{figure*}

Fig.~\ref{fig:Fig1}a illustrates the survey acquired for the two different surface conditions: as-cleaved (blue curve) and aged, \textit{i.e.}, after 72 hours in HV (red curve). We can see that the intensities of the signals from I 3\textit{d} and 4\textit{d} edges (Fig.~\ref{fig:Fig1}b) and V 2\textit{p} edges (Fig.~\ref{fig:Fig1}c) -- the former in particular -- are strongly reduced in the aged sample; additionally, strong C 1\textit{s} and O 1\textit{s} peaks appear \textcolor{black}{(Fig. \ref{fig:Fig1}a)}, suggesting a contaminated surface environment. Further examination of the element-specific photoemission edges give more information on the chemical condition of the surface in the two samples. On one hand the I 3\textit{d} and 4\textit{d} signals, albeit much less intense, show an effectively unaltered lineshape (Fig.~\ref{fig:Fig1}b); on the other hand, the V 2\textit{p} doublet displays additional features at higher binding energies (red arrows in Fig.~\ref{fig:Fig1}c). The above suggests that the chemical environment of vanadium is considerably altered at the surface of VI\textsubscript{3}, whereas the chemical state of bonded iodine is mostly unchanged. A structured O 1\textit{s} peak (black arrow in Fig.~\ref{fig:Fig1}c) is also present on the aged sample and hints at a complex chemical state of this element at the surface. Since I core levels are not modified by spurious oxygen contamination, we consider the O 1\textit{s} peak as composed by a combination of V-O bonds and molecular oxygen, whereas V 2\textit{p} combines stoichiometric VI\textsubscript{3} features, here almost vanished, with the formation of \textcolor{black}{other high-valence V compounds}. A more detailed analysis will be given in the following. 

\begin{figure}[htb]
\includegraphics[width=0.95\linewidth]{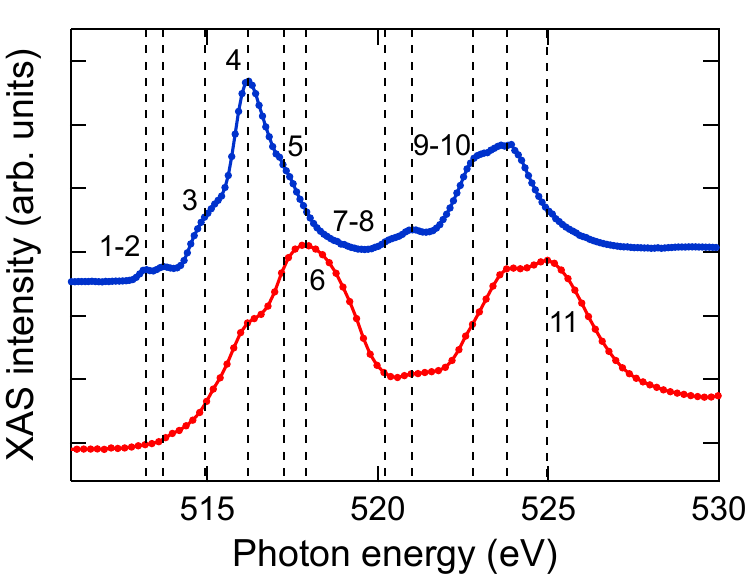}
\caption{\protect\label{fig:Fig2} XAS spectra at V \textit{L}\textsubscript{2,3} edges on as-cleaved (blue) and aged (red) samples. The photon flux for these spectra was equivalent to $\SI{2.8e9}{\photons\per\s}$ at the gold-plated mesh grid before the endstation.}
\end{figure}

To capture more details of the valence state of V and further understand the role of the metallic centres in the chemical modification on the surface, we measured XAS spectra across the V \textit{L}\textsubscript{2,3} edges for both samples. Experimental spectra are shown in Fig.~\ref{fig:Fig2}. The lineshape of the as-cleaved spectrum follows expectations for a V\textsuperscript{3+} oxidation state, including all the pre-edge features (1-2) and (7-8) from empty \textit{t\textsubscript{2g}} states \cite{DeVita2022,Sant2023,DeVita2024}. Conversely, the aged sample presents an overall shift towards higher photon energies, as well as significant differences in lineshape. The main \textit{L}\textsubscript{3} peak (6) displays a shift of 1.6 eV compared to pristine VI\textsubscript{3} (4) and an overall skew towards high-energy components of the multiplet; the lineshape and bandwidth of the \textit{L}\textsubscript{2} edge is similarly altered; we also note the absence of well-defined pre-edge states.

The results suggest a relevant change in the orbital configuration of V upon surface contamination. In particular, the lineshape of the aged sample displays similarities with 4+ and 5+ valence states, by comparison with absorption spectra of vanadates and vanadium oxides \cite{Abbate1993,Zimmermann1998,Wu2018,Polewczyk2023b,DElia2024}. It is likely that this modification is significantly affecting more than the first atomic layer, since the probing depth of the technique amounts to \SIrange{4}{6}{\nano\metre} in the employed photon energy range \cite{Frazer2003}: this corresponds to around seven unit cells in the monoclinic phase, each composed of a double layer of VI\textsubscript{6} octahedra (space group $C2/m$, $c=\SI{6.9502\pm0.0004}{\angstrom}$ \cite{Tian2019}).

\begin{figure*}[htb]
\includegraphics[width=0.9\linewidth]{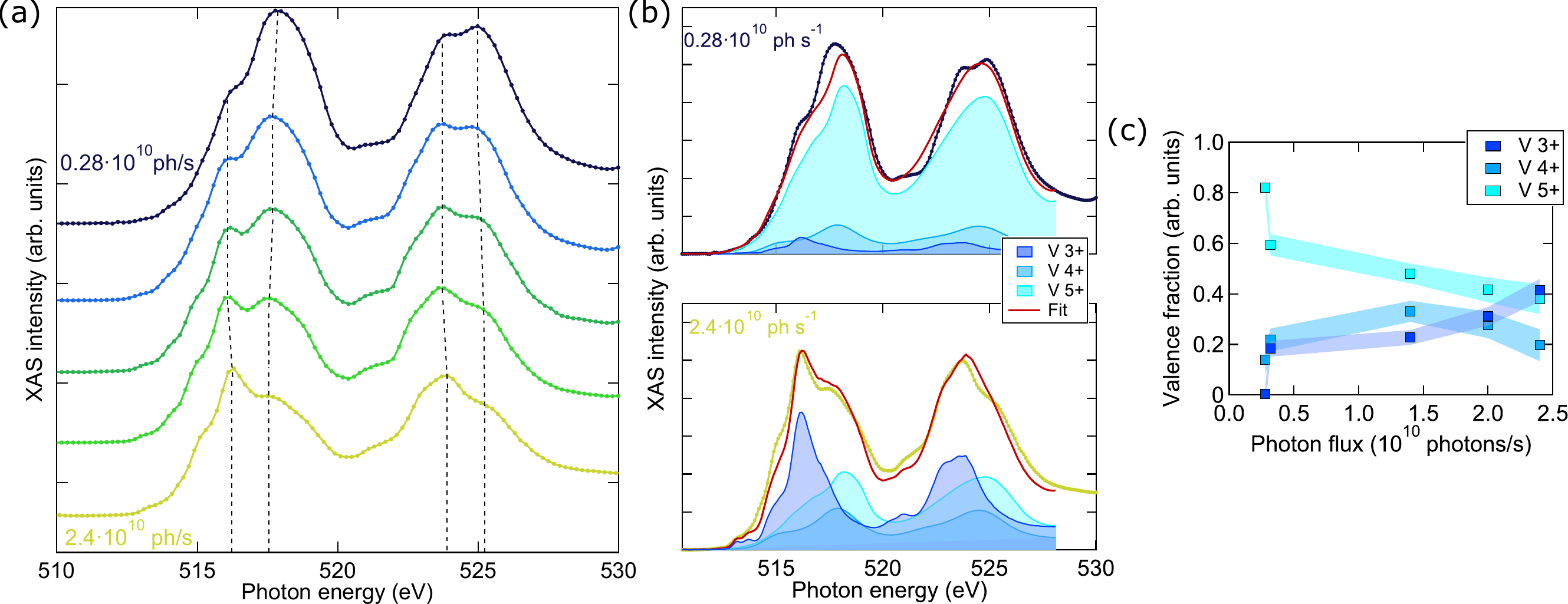}
\caption{\protect\label{fig:Fig3} (a) Subsequent XAS \textcolor{black}{spectra} at V \textit{L}\textsubscript{2,3} edges on the same probed area, from top to bottom, under increasing photon flux irradiation. (b) Fitted XAS spectra with reference spectra for the first spectrum at low irradiation flux (top black curve) and last measure at high irradiation flux (bottom yellow curve). (c) Evolution of the valence fraction contribution of the measured spectra as a function of irradiation flux.}
\end{figure*}

\textcolor{black}{We note that no significant evolution takes place on the pristine VI\textsubscript{3} surface when low-flux X-rays ($\approx\SI{0.28e10}{\photons\per\s}$ on the gold-plated mesh grid) have been continuously shone on the sample for several hours (see SI Fig.~S1). Therefore, the electronic properties of as-cleaved VI\textsubscript{3} are unaltered by low-flux X-ray illumination.} On the other hand, once the photon flux was increased up to $\approx\SI{2.40e10}{\photons\per\s}$, the lineshape of the XAS spectra on the aged sample changed dramatically, as displayed in Fig.~\ref{fig:Fig3}a. In particular, we notice a gradual increase of low-energy features of the spectrum, as well as, importantly, the appearance of pre-edge signals. This change is not a transient effect, as decreasing the flux does not result in a recovery of the high-energy features. Conversely, when probing another spot on the sample surface, the previous lineshape is retrieved (SI, Fig.~S2). These are indicators of a local effect of soft X-ray irradiation on the exposed area of the \textcolor{black}{partially hydrated} \textcolor{black}{crystal surface}, which amounts to a chemical modification of the surface bonds between the crystal and adsorbates.

In order to quantify the change in the chemical state of V as probed by XAS, we fitted the experimental spectra at different flux intensities with a linear combination of the spectral features ascribed to reference V\textsuperscript{x+} spectra, as measured on the same beamline. The reference spectra belong respectively to pristine bulk VI\textsubscript{3} (V\textsuperscript{3+}), V\textcolor{black}{O}\textsubscript{2} (V\textsuperscript{4+}), and the over-\textcolor{black}{oxidized} surface layer of SrVO\textsubscript{3} (V\textsuperscript{5+}) \cite{DeVita2022,Polewczyk2023,DElia2024}. In the case of V\textsuperscript{4+} and V\textsuperscript{5+} contributions, a broadening has been included to account for the absence of \textcolor{black}{single-crystal} order of these compounds on the VI\textsubscript{3} surface and the consequent blurring of sharp multiplet peaks. The procedure is performed on a range up to \SI{528}{\eV}, due to the interference of the overlapping O edge especially for the V\textsuperscript{5+} component. Our fits succeed in reproducing the main features of the measurements (Fig.~\ref{fig:Fig3}b-c) and reveal that the main contribution at low flux comes from \textcolor{black}{V\textsuperscript{5+} states, with a minority contribution from V\textsuperscript{4+}. The preminence of a V\textsuperscript{5+} oxidation state is consistent with the results in ref. \cite{Mastrippolito2022} , where a major V\textsuperscript{5+} and a minor V\textsuperscript{4+} components were revealed by XPS on air-exposed VI\textsubscript{3} crystals. The quantitative differences can be ascribed to the procedural standards applied in the two cases, mainly regarding the time of exposure, and the composition and background pressure of the atmospheric contaminants. After irradiation, the stoichiometric V\textsuperscript{3+} component is strongly enhanced. As seen in Fig.~\ref{fig:Fig3}c, the V\textsuperscript{5+} contribution shows a monotonic decrease as a function of flux, whereas the V\textsuperscript{4+} component initially increases. We can tentatively ascribe such evolution upon x-ray irradiation to beam-induced photoreduction of high-valence vanadates at low photon flux.} 

\textcolor{black}{Importantly, compared to ref. \cite{Mastrippolito2022} we reveal a V\textsuperscript{3+} signal on our aged samples, unreported in their measurements. This is likely a consequence of the much lower exposure to contaminants of the aged sample, which partly retains a portion of stoichiometric VI\textsubscript{3} within the probed area. Since no V\textsuperscript{3+} signal was found on air-exposed samples in ref.\cite{Mastrippolito2022} , we deduce that} \textcolor{black}{vanadates} \textcolor{black}{with 3+ valence character are unlikely to form from VI\textsubscript{3} degradation. This justifies our choice to use only the VI\textsubscript{3} lineshape retrieved from the as-cleaved sample to fit the V\textsuperscript{3+} component.}

\begin{figure}[htb]
\includegraphics[width=0.9\linewidth]{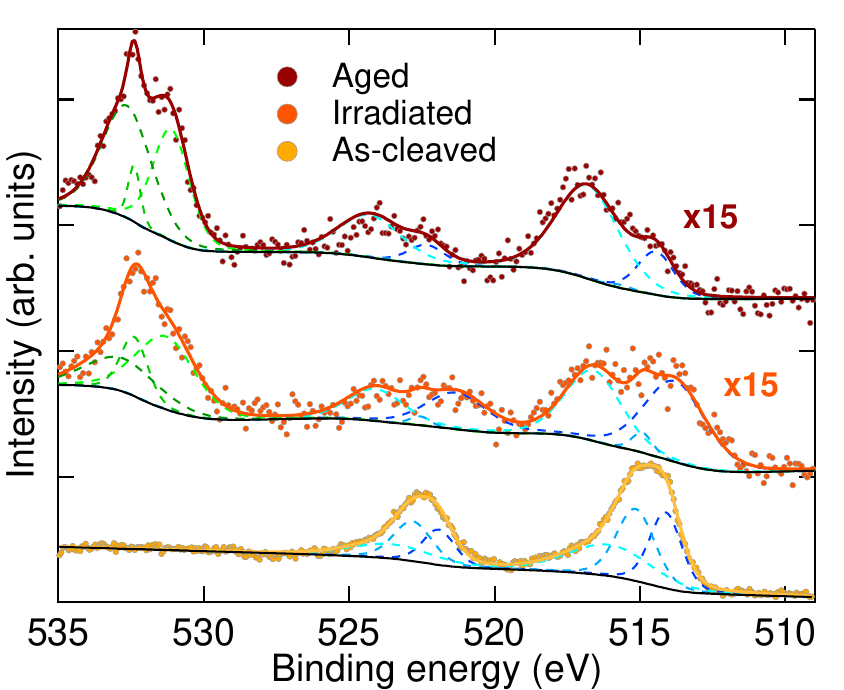}
\caption{\protect\label{fig:Fig4} XPS spectra of as-cleaved, aged and irradiated samples; solid lines are fits of the experimental curves, dashed lines are the different components for V (blue/teal) and O (green) signals.}
\end{figure}

To evaluate the effect of X-ray irradiation \textcolor{black}{on the VI\textsubscript{3} surface} when comparing clean and aged surfaces, we also carried out further XPS measurements of the V 2\textit{p} core level and the adjacent O 1\textit{s} peak, on as-cleaved, aged and irradiated samples (Fig.~\ref{fig:Fig4}). Here, "irradiated" refers to the surface area that was exposed to the maximum X-ray flux and consequently displayed the changes in the XAS spectrum for a photon flux $\approx\SI{2.40e10}{\photons\per\s}$ in Fig.~\ref{fig:Fig3}. XPS peaks were fitted with multiplet peaks to disentangle the different contributions. \textcolor{black}{We note that, since the probing depth of XAS and XPS are considerably different, a direct, quantitative comparison between the two would be a daunting task; however, XPS results are complementary to the} XAS analysis to understand the relation between the valence, chemical state, and degree and type of contamination at the surface of the crystal. 

In the XPS spectra, we can see how the aged sample shows the majority of the V 2\textit{p} signal coming from higher binding energies compared to the as-cleaved one. After irradiation, the component at lower binding energies has emerged, consistently with what seen in XAS (Fig.~\ref{fig:Fig3}c). The O 1\textit{s} peak contains several components, corresponding to different bound states. At the lowest binding energy, the O signal marks the presence of vanadium-O\textsuperscript{2-} bonds; the higher binding energy component can be likely ascribed to \textcolor{black}{O-H bonds due to} surface bonded H\textsubscript{2}O replacing the I\textsuperscript{-} ligands \cite{Linn1984,Choudhury1989,Mastrippolito2022}, with possibly a minor contribution from O-C bonds \cite{Ren2020}. The element ratio O:V in mol\% can be approximately estimated as 1.7:1 in the aged sample, and 1.3:1 in the irradiated sample, by combining the photoionization cross sections and the respective integrated areas. It has to be underlined that this rough estimation does not take into account the different distribution of the two atomic species within the probed thickness. Given that XAS, a technique \textcolor{black}{whose probing depth amounts from \SIrange{4}{6}{\nano\m}\cite{Frazer2003}}, also reveals chemical changes upon irradiation, we consider reasonable the assumption that, within the probing depth of XPS at these photon energies (around \SI{1}{\nano\metre} \cite{Seah1979}), both O and V are evenly distributed in the \textcolor{black}{probed area}. The overall O 1\textit{s} signal in Fig. \ref{fig:Fig4} decreases by roughly \SI{25}{\percent} after irradiation; within the multiplet structure, 
\textcolor{black}{the peak due to O-H bonds, \textit{i.e.} water molecules, }is more strongly suppressed, whereas the peak \textcolor{black}{due to V-O\textsuperscript{2-} bonds} 
decreases less dramatically and considerably broadens. Conversely, the integrated areas of the V 2\textit{p} structures do not change within the level of confidence of the fit. Compared to the expectations for compounds displaying 4+ or 5+ valence, the relatively low O:V ratio accounts also for the presence of clean VI\textsubscript{3}.

From these observations we can surmise that the impinging X-rays at mild photon flux intensities partially desorb oxygen and H\textsubscript{2}O molecules hydrating VI\textsubscript{3}, and change the coordination and bonding of the vanadium oxides. \textcolor{black}{We also consider that, in principle, we cannot exclude also a beam-induced photoreduction process, \textit{i.e.}, the formation of 3+ valence oxides giving rise to the increase of the V 2\textit{p} signal at low binding energies.} \textcolor{black}{As a matter of fact,} \textcolor{black}{we note that the fraction of V signal at low binding energies increases after irradiation,} \textcolor{black}{whereas} \textcolor{black}{the oxygen-to-vanadium ratio is still relatively high, suggesting that oxides are still present. However,} the appearance of pre-edge features at V \textit{L}\textsubscript{3} edge of XAS spectra after high flux exposure (see Fig.~\ref{fig:Fig3}c) suggests \textcolor{black}{that the 3+ signal is mainly attributable to} stoichiometric pristine VI\textsubscript{3}. \textcolor{black}{A tentative interpretation could be that partial desorption of oxygen and H\textsubscript{2}O exposes more underlying VI\textsubscript{3}, causing an increase in the V\textsuperscript{3+} signal. Further support to this explanation is given by an increase of the photoemitted intensity upon irradiation of the aged sample on the I 3\textit{d} and 4\textit{d} core levels measured in XPS, with no changes in energy position or lineshape, \textit{i.e.} no change in the coordination of the iodine (Supporting Information, Fig.~S5). However, a full explanation would require a more thorough and statistically significant investigation of the core level peaks as a function of irradiation}. Given the likely high degree of surface disorder and the consequent lower statistics in the acquired spectra, assigning several multiplet peaks as in the as-cleaved sample is a much more challenging task for the aged and irradiated cases. In any case, the as-cleaved sample itself shows a nontrivial multiplet structure, which cannot be fitted by a single 2\textit{p} doublet. The lack of a well-defined XPS peak suggests that the electronic state of VI\textsubscript{3} might be more complicated to understand than a simple V\textsuperscript{3+} state. For example, Mastrippolito \textit{et al.} have suggested that even pristine VI\textsubscript{3} intrinsically hosts trapped electrons giving rise to V\textsuperscript{2+} localized states \cite{Mastrippolito2023}. Moreover, there is a growing consensus that a mixed-orbital state characterized by strong covalency of the metal-ligand bond is instrumental in correctly describing the electronic and magnetic properties of transition metal halides \cite{Sant2023,He2025,DeVita2025}, making the interpretation of XPS spectra more challenging. \textcolor{black}{Therefore, a complete understanding of the microscopic mechanisms taking place at the contaminated VI\textsubscript{3} surface upon irradiation is beyond the scope of this work.}

\section{Conclusions}

In this study, we combined X-ray absorption and X-ray photoemission spectroscopies to investigate the chemical stability of freshly cleaved VI\textsubscript{3} single crystals. We observed how surface degradation induced by the interaction of the hygroscopic surface with spurious oxygen and water molecules, even in UHV conditions, leads to chemical alteration of the first monolayers, with the formation of vanadates. \textcolor{black}{We investigated the effects of soft X-ray irradiation under mild photon fluxes of the contaminated surfaces} \textcolor{black}{of crystals aged in HV}, \textcolor{black}{unveiling the induced valence modifications on V and I core levels}.

These results provide experimental evidence of the relatively weak interaction between \textcolor{black}{contaminants and VI\textsubscript{3} surface layers. We prove that} low-energy means \textcolor{black}{can locally alter the chemical environment of the different surface molecular species, and partially remove contamination, with deep implications in the search for possible routes for functionalization of VI\textsubscript{3} surface chemical properties. However, this effect is not necessarily benign, and further studies are required to understand if a full restoration of the pristine surface} \textcolor{black}{ can actually} \textcolor{black}{take place, or whether oxide reduction plays an important role in the formation of V\textsuperscript{3+} oxides on the surface.}

\subsubsection{Supporting Information} 

VI\textsubscript{3} crystal structure, RGA information of the UHV system, XPS surveys to monitor the as-cleaved surface evolution, XAS spectra as a function of position on the irradiated sample, XPS spectra of I 3\textit{d} and 4\textit{d} core levels on aged and irradiated surfaces.

\section{Acknowledgments}
A.D.V. acknowledges financial support from the Max Planck Society and BERLIN QUANTUM, an initiative endowed by the Innovation Promotion Fund of the city of Berlin. We thank A. Fondacaro for the support in devising and commissioning the glovebox-UHV handling and transfer system. This work was performed in the framework of the NFFA-SPRINT facility, supported by MUR as the Activity of International Relevance NFFA (www.trieste.NFFA.eu). G.P. acknowledges financial support from PNRR MUR project PE0000023-NQSTI. We thank G. Fratesi, S. Achilli \textcolor{black}{and L. Braglia }for fruitful discussion.

\section*{Author Contributions}

A.D.V. and G.P. conceived the research project. A.D.V., V.P., and G.V. performed the synchrotron measurements. A.D.V. analyzed the data under the supervision of G.V. A.D.V. and G.V. wrote the manuscript, with contributions from all authors.



\bibliography{bibliography}

\end{document}




\title{Supporting Information - Valence Modifications in Hygroscopic VI\textsubscript{3} Degraded Crystals via Soft X-Ray Synchrotron Radiation}

\author{A. De Vita}
\email{alessandro.de.vita@tu-berlin.de}%
\affiliation{Institut für Physik und Astronomie, Technische Universität Berlin, Strasse des 17 Juni 135, 10623 Berlin, Germany\looseness=-1}%
\affiliation{Fritz Haber Institute of the Max Planck Society, Faradayweg 4--6, 14195 Berlin, Germany\looseness=-1}%
\author{V. Polewczyk}
\affiliation{Université Paris-Saclay, UVSQ, CNRS, GEMaC, 78000, Versailles, France\looseness=-1}%
\author{G. Panaccione}%
\affiliation{CNR - Istituto Officina dei Materiali (IOM),  S.S. 14, km 163.5, 34149 Trieste, Italy\looseness=-1}%
\author{G. Vinai}
\email{vinai@iom.cnr.it}
\affiliation{CNR - Istituto Officina dei Materiali (IOM),  S.S. 14, km 163.5, 34149 Trieste, Italy\looseness=-1}%

\maketitle
\thispagestyle{fancy}

\section*{Additional Figures}

For completeness, we present in Fig.~\ref{fig:SI1} the XPS survey at \SI{800}{\eV} acquired on pristine, clean VI\textsubscript{3} right after cleaving, after four hours, and after eight hours. The photon flux was quantified as \SI{0.28e10}{\photons\per\s}. We do not find any evidence of degradation or modifications of the core levels.

\renewcommand{\thefigure}{S1}
\begin{figure}[htb]
\centering
\includegraphics[width=0.5\linewidth]{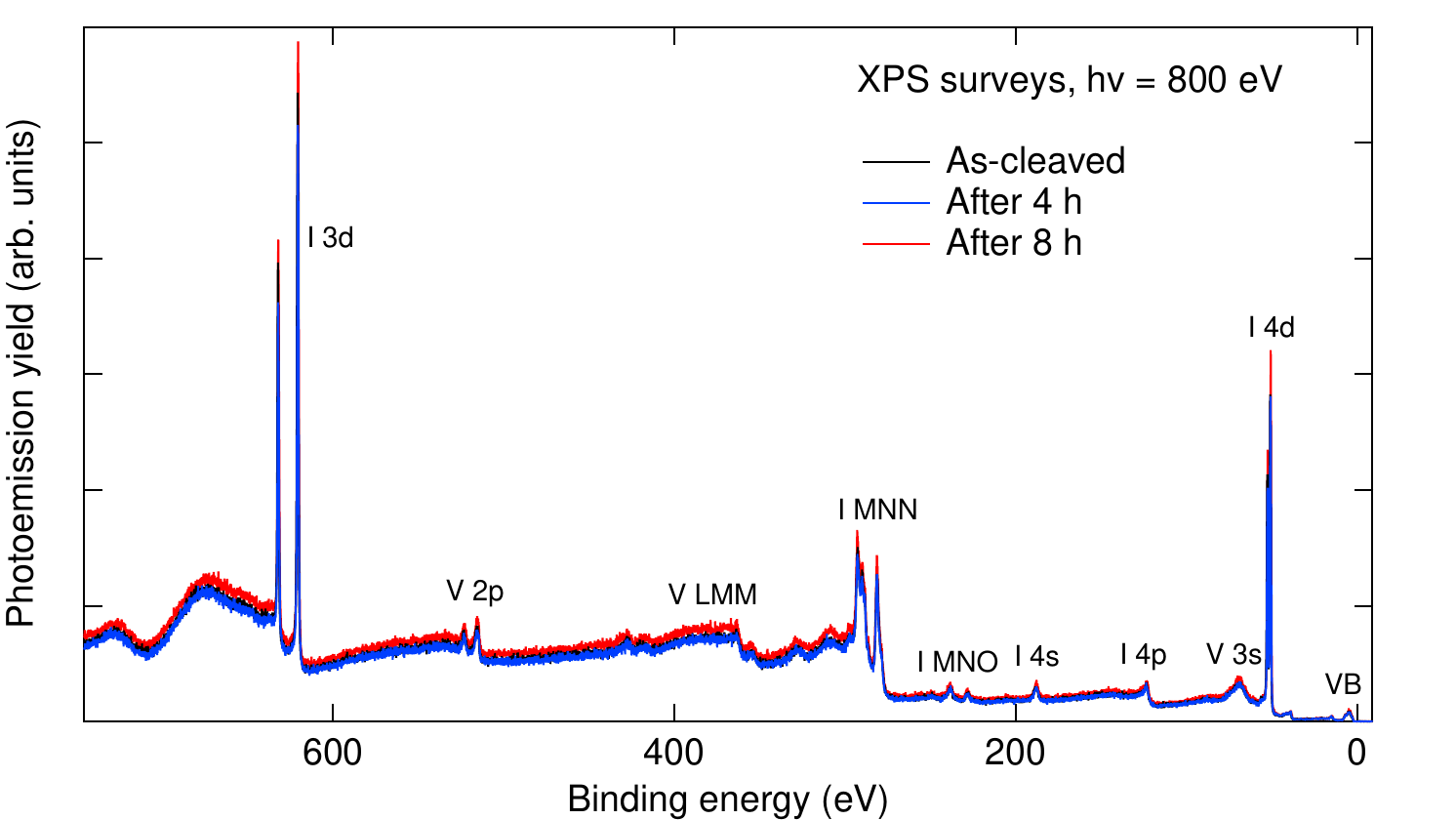}
\caption[captionsep=none]{\protect\label{fig:SI1}XPS surveys acquired at \SI{100}{\K} with \SI{800}{\eV} photons.}
\end{figure}

In Fig.~\ref{fig:SI3} we show the effect of irradiation on the measured spot compared to other areas on the sample surface. In Fig.~\ref{fig:SI3}a, the 2D map of the sample displays three spots (A, B, and C) where data have been acquired. Spot A has been irradiated with the maximum photon flux ($\SI{2.8e9}{\photons\per\s}$, Fig.~3a of the main text), spot B and C are pristine. Fig.~\ref{fig:SI3}b collects the XAS spectra across the V \textit{L}\textsubscript{2,3} edges in each of the three spots -- the top and bottom curves (A-1 and A-2) have both been performed on A, respectively before and after the acquisition of the spectra B and C. The photon flux was $\SI{2.8e9}{\photons\per\s}$. It is clear from the lineshape and the position of the main peaks that, after exposure to the maximum flux, the irradiated spot is characterized by a different chemical state with respect to the other areas of the sample, and that such effect is not transiently appearing under the beam.

\renewcommand{\thefigure}{S2}
\begin{figure}[htb]
\centering
\includegraphics[width=0.95\linewidth]{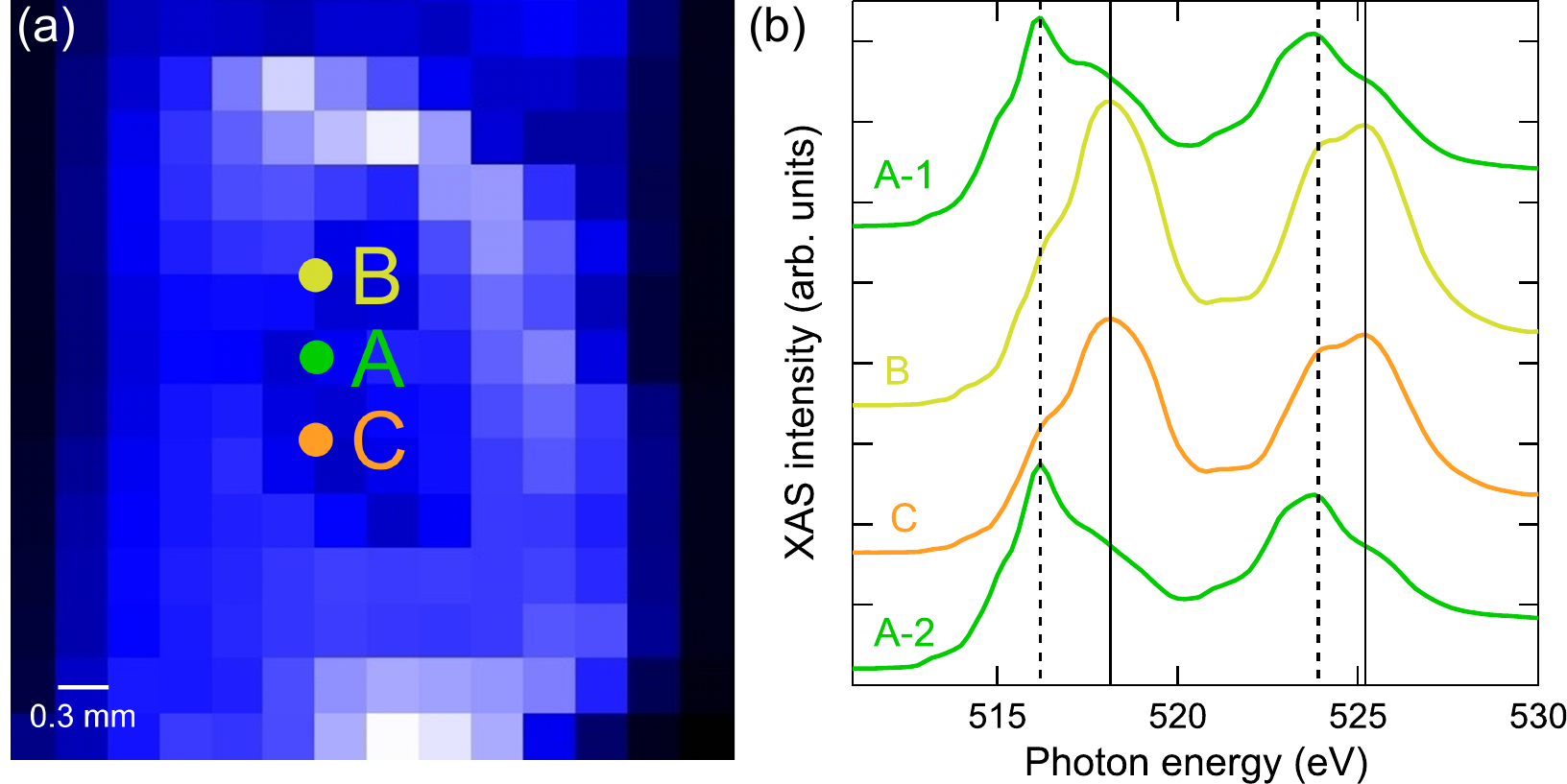}
\caption{\protect\label{fig:SI3}(a) 2D map of the sample holder area; the color scale is the TEY signal (lighter corresponds to higher signal intensity). The bright pixels indicate the edges of the sample. (b) XAS spectra across the V \textit{L}\textsubscript{2,3} edges for a photon flux $\SI{2.8e9}{\photons\per\s}$; the colors correspond to the spots in (a) where the spectrum was measured. The labels A-1 and A-2 refer both to spectra measured on A at different times, as explained in the main text. The dashed and solid lines highlight the main \textit{L}\textsubscript{3} and \textit{L}\textsubscript{2} peaks for the irradiated and aged surfaces, respectively.}
\end{figure}

\renewcommand{\thefigure}{S3}
\begin{figure}[htb]
\centering
\includegraphics[width=0.55\linewidth]{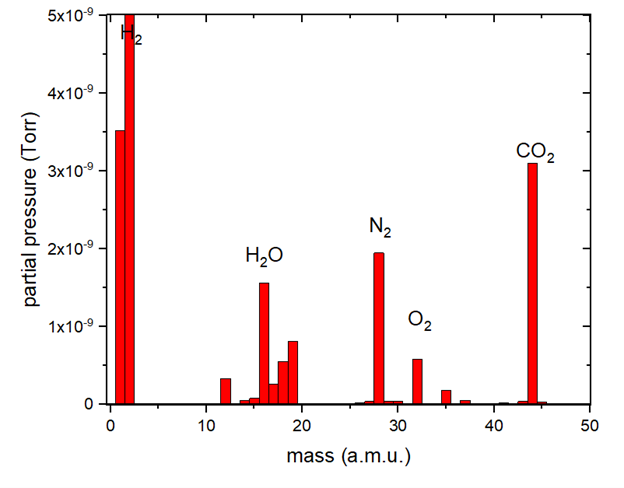}
\caption{\protect\label{fig:SI4}Residual Gas Analysis (RGA) for the vacuum system where the aged sample has been exposed to surface contamination.}
\end{figure}

In Fig.~\ref{fig:SI4} we show Residual Gas Analysis (RGA) information, directly measured in the same experimental chamber where the samples were exposed to background pressure for 72 hours, in order to provide a more quantitative assessment of the composition of residual gases. In particular, the spectrum shows that the largest percentages of partial pressure are shared between H\textsubscript{2}, water vapor (H\textsubscript{2}O), N\textsubscript{2}, and carbon-derived species (CO/CO\textsubscript{2}). Indeed, in our XPS surveys (Fig.~1a of the main text) we reveal sizeable fingerprints of O and C on the sample that has been exposed to \SI{2600}{\langmuir}, due to the adsorbed species. At the photon energies we employed, the photoionization cross section of the H 1\textit{s} is too small for detection at our beamline (several orders of magnitude smaller than those of O, C or V); therefore, no signature of H\textsubscript{2} is revealed. 

\renewcommand{\thefigure}{S4}
\begin{figure}[htb]
\centering
\includegraphics[width=0.95\linewidth]{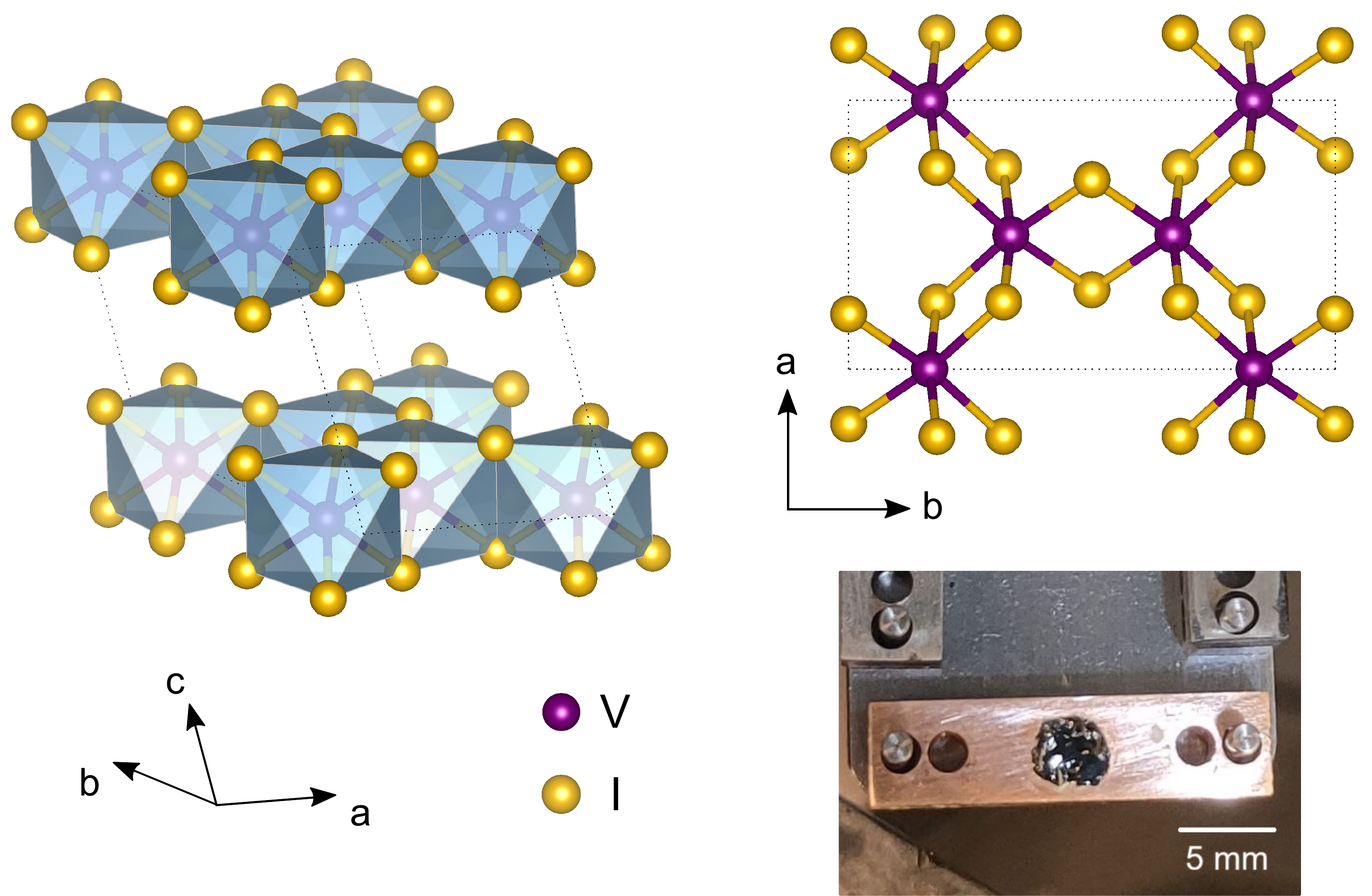}
\caption{\protect\label{fig:SI5}Views of the crystal structure of VI\textsubscript{3}, from the side and along the \textit{c} axis. \textit{Bottom right}: photo of a VI\textsubscript{3} crystal as mounted on a Cu sample holder.}
\end{figure}

\textcolor{black}{In Fig.~\ref{fig:SI6} the XPS spectra of I 3\textit{d} and 4\textit{d} core levels, and the corresponding spin-orbit split doublet fitting, are shown. The spectra have been measured on the sample in aged and irradiated conditions. The extracted values of spin-orbit splitting (\SI{11.5}{\eV} for I 3\textit{d}, \SI{1.7}{\eV} for I 4\textit{d}) are consistent with the expected values. It is clear that the photoemission intensity increases upon irradiation: quantitative analysis of the I:V ratio in mol\unit{\percent} (using the V 2\textit{p} in Fig.~4 of the main text as reference) reveals that the area of the photoemission signal, and the corresponding I:V ratio, increases by $\approx\SI{60}{\percent}$ at the 3\textit{d} peaks after irradiation, and by $\approx\SI{35}{\percent}$ at the I 4\textit{d} peaks. Electrons photoemitted from I 3\textit{d} peaks have $\sim\SI{150}{\eV}$ kinetic energy, and thus their inelastic mean free path is roughly half that of electrons from I 4\textit{d} peaks, which have $\sim\SI{750}{\eV}$ kinetic energy and therefore are more bulk-sensitive. This offers further support to the hypothesis: due to the higher surface sensitivity of the I 3\textit{d} signal, their peaks are more affected by the desorption of surface contaminants and the exposure of the stoichiometric VI\textsubscript{3} underneath.}

\textcolor{black}{We also note that no change in the energy position of the peaks, nor in their lineshape, takes place upon irradiation. This indicates that iodine undergoes no significant change in its coordination or chemical environment. It is thus likely that, even upon mild contamination by H\textsubscript{2}O, iodine migrates/desorbs from the surface, and the remaining photoemission signal comes from atoms buried in lower layers, unaffected by moisture and thus still coordinated with vanadium. This would justify the high chemical stability of iodine upon irradiation.}

\renewcommand{\thefigure}{S5}
\begin{figure}[htb]
\centering
\includegraphics[width=0.95\linewidth]{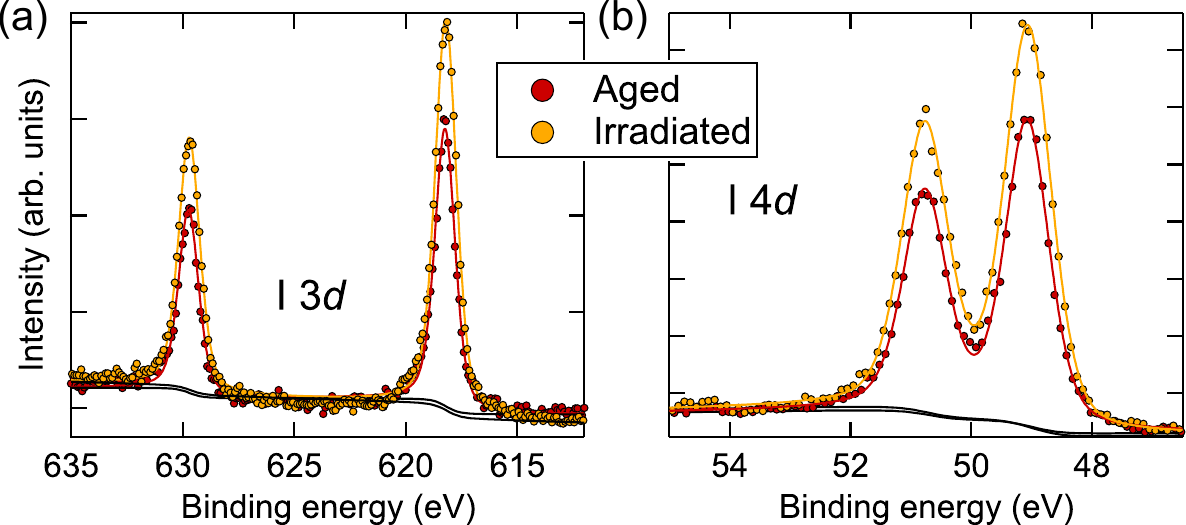}
\caption{\protect\label{fig:SI6}XPS spectra of the I 3\textit{d} (a) and 4\textit{d} (b) core levels. Solid colored lines are fits of the experimental curves, solid black lines are the superimposed Shirley background.}
\end{figure}

